\documentclass[twocolumn,amssymb, amsmath,nobibnotes, aps, prb]{revtex4}
\usepackage[dvips]{graphicx}
\usepackage{color}

\begin{document}

\title{Pressure Induced Change in the Magnetic Modulation of CeRhIn$_5$}
\author{S. Majumdar}
\email{phshd@warwick.ac.uk}
\author{G. Balakrishnan}
\author{M. R. Lees}
\author{D. M$^c$K. Paul}
\affiliation{Department of Physics, University of Warwick, Coventry CV4
7AL, UK}

\author{G. J. McIntyre}

\affiliation{Institut Laue-Langevin, B.P. 156, 38042 Grenoble Cedex 9,
France}

\date{\today}

\begin{abstract}

We report  the  results of a  high pressure neutron diffraction study of
the heavy fermion compound CeRhIn$_5$ down to 1.8 K. CeRhIn$_5$ is known
to order magnetically below 3.8 K with an incommensurate  structure. The
application of  hydrostatic pressure up to 8.6 kbar  produces no change in
the magnetic wave vector {\bf q}$_m$. At  10 kbar of pressure however, a
sudden change in the magnetic structure occurs. Although the magnetic
transition temperature remains the same, {\bf q}$_m$ increases from (0.5,
0.5, 0.298) to (0.5, 0.5, 0.396). This  change in the magnetic modulation
may be the outcome  of a change in the electronic character of this
material  at 10 kbar.

\end{abstract}

\maketitle

\par There is considerable evidence that in heavy fermion systems the
existence of  {\it partially Kondo compensated low moment magnetic order}
is associated with the occurrence of  pressure-induced superconductivity.
A wealth of Ce based compounds with a magnetically ordered ground state,
such as CeCu$_2$Ge$_2$, CePd$_2$Si$_2$, CeIn$_3$
~\cite{cerh2si2,jaccard,grosche,walker, mathur},  have been found to show
superconductivity under an applied hydrostatic pressure. The application of
pressure gradually  destroys the magnetic order of the system  and
eventually a superconducting state emerges byeond the critical pressure $P_c$. In  contrast to normal BCS superconductors, it is believed that  the magnetic spin fluctuations
rather than the lattice vibrations effectively bind the electrons (more
precisely the  heavy electrons) into Cooper pairs~\cite{grosche1}. Many of
these compounds also show non-Fermi liquid behavior in  the vicinity of
the magnetic-nonmagnetic phase boundary~\cite{nfl}. In the last  few years
there has been considerable experimental as well as theoretical effort to
understand the possible role of magnetic interactions in creating  a
superconducting ground state~\cite{monthoux, mathur}.
 
\par
The recently discovered heavy fermion compound CeRhIn$_5$ provides a useful opportunity
to investigate the coupling between magnetism and superconductivity
because of its reasonably high transition temperature ($T_c$ $\sim$ 2 K)
in an accessible pressure (16 kbar) range. This material is also different in that there appears to be a sharp boundary where superconductivity develops then exists over an extended pressure range with little variation  in  T$_C$.

\par
CeRhIn$_5$ crystallizes in a tetragonal HoCoGa$_5$ structure  with
alternating layers of  CeIn$_3$  and RhIn$_2$  arranged along the
$c$-axis.  At ambient pressure  it orders magnetically at T$_N$ =  3.8 K
into a $c$-modulated incommensurate structure~\cite{hegger}.  Neutron
diffraction~\cite{neutron1bar} and nuclear quadrapole resonance~\cite{nqr}
experiments  indicate that below $T_N$  the magnetic  moment at each Ce site is  0.26
$\mu_B$ with the spins lying in the basal plane. In the
basal plane, the magnetic moments  adopt a nearest neighbor
antiferromagnetic  structure, while along  the $c$-axis they form an
incommensurate spiral with  a pitch of $\delta$ = 0.298, corresponding to
a turn angle of 107$^{o}$ between  the successive CeIn$_3$ layers. The
resulting magnetic modulation vector is  {\bf q}$_m$ = (0.5, 0.5, 0.298).
The  application of  hydrostatic pressure up to  9 kbar has little effect
on the value of $T_N$. Beyond 16 kbar of pressure, resistivity  and heat
capacity  data suggest  that  CeRhIn$_5$  undergoes a first order like
transition to  a superconducting ground state (T$_c \sim $ 2 K)  with the
sudden disappearance of the 3.8 K  magnetic phase~\cite{hegger}.
Resistivity {\it vs.} temperature  data~\cite{hegger} indicate a small
sharp drop around the temperature $T_? \approx$ 3 K  above 9 kbar of
pressure,  which continues to exist in the superconducting state (above 16
kbar of pressure). The signature of the  long range magnetic order in the
the heat capacity {\it vs.} temperature data above 3 kbar is indicated by
a  rather broad maximum ($T_{max}$), which is close to the ambient
pressure value of $T_N$ up to 9 kbar. Although, heat capacity
measurements~\cite{heatcap} fail to resolve $T_?$ and $T_N$ in the
pressure  range  9 to 15 kbar, it indicate a gradual shift of $T_N$ (as
indicated by $T_{max}$) to lower temperatures above 10 kbar of pressure.
Above  $P_c$, a very  broad anomaly around 3 K, presumably of magnetic
origin,  continue to exists in the heat capacity {\it vs.} temperature
data, which matches closely with the value of $T_?$ obtained from the
resistivity data. 
\par
The interpretation of both resistivity and heat capacity experiments as to the onset of long range order is not straight forward. In the case of resistivity, the very small change observed at $T_?$ could result from many causes. For heat capacity data there is a smooth evolution in the shape of the peaked structure, which is clearly associated with T$_N$ at atmospheric pressure but may have other origins at higher applied pressure. Neutron experiments definitively map the onset of an ordered moment through the appearance of Bragg reflections from the magnetic structure.

\par
The temperature-pressure phase diagrams of  Ce based heavy fermion
superconductors~\cite{mathur} such as CeIn$_3$ and  CePd$_2$Si$_2$ show a
gradual decrease of $T_N$  with increasing pressure before the onset of
superconductivity.  In  case of
CeRhIn$_5$, $T_N$ is quite robust with respect to the application
pressure up to 10 kbar. The superconductivity also occurs with a sudden
disappearance of $T_N$ at $P_c$.
\par
It is therefore essential to understand  the nature of the magnetic
phase of CeRhIn$_5$  which is stable at high pressure. Bao {\it et
al.}~\cite{neutron3p8kbar} reported  neutron diffraction data of
CeRhIn$_5$ at 3.8 kbar  They observed a magnetic modulation vector of
{\bf q}$_m$ = (0.5, 0.5, 0.294), which is almost identical to the ambient
pressure value. In this paper we report the results of a high pressure (up
to 13 kbar)  neutron diffraction study of the magnetic ordering of
CeRhIn$_5$. Our investigation shows no variation in the magnetic structure
up to a pressure of 8.6 kbar. Above 10 kbar  of pressure we observed
a  marked change in {\bf q}$_m$ from (0.5, 0.5, 0.298) to (0.5, 0.5, 0.396).  Despite
this change in the {\bf q}$_m$, $T_N$  remained constant up to
11 kbar of pressure. At 11 kbar, no anomaly was detected in the magnetic
reflections in the vicinity of $T_?$ = 3 K, the temperature at which an
unidentified feature was seen in the resistivity data~\cite{hegger}.
Surprisingly, at 13 kbar  we failed to observe any magnetic reflections
along the (0.5, 0.5, $\ell$) direction for temperatures above our base of 1.8 K.

\par
Single crystals of CeRhIn$_5$ were grown using the indium flux
technique~\cite{xtal}. For this experiment, a bar shaped (1.5 $\times$
1.5$\times$ 4 mm$^3$) crystal was cut from a large single crystal. The
crystal was mounted with the [110] direction  vertical. This enabled us to
reduce the absorption of neutrons by the indium nuclei for our diffraction
scans along the direction of the reciprocal lattice vector (0.5, 0.5,
$\ell$) ($\ell =$ 0 to 1). The high pressure neutron diffraction
experiment was carried out using  the  thermal-neutron diffractometer D10,
at the Institut Laue-Langevin, Grenoble, France, in the two-axis mode to
accomodate the large He-flow cryostat. A  clamp-type alumina pressure cell
(inner diameter  6 mm and outer diameter 7 mm)  was used  for the
generation of hydrostatic pressure with fluorinert as the pressure
transmitting medium. The  monochromatic neutron beam of wavelength 2.359
\AA~ used for the diffraction experiment  was  obtained by reflection from
a pyrolytic graphite monochromator.  In order to reduce the background
counts from the pressure cell and the cryostat, a pyrolytic graphite
analyzer was included for most of the scans.  A sodium chloride crystal was
placed alongside  the sample in the pressure cell.  Its lattice parameters
were determined independently,  which  enabled us to measure the actual
pressure  in  the cell {\it in situ}  by comparing  the known pressure
dependence of the lattice parameter of sodium chloride with these
measurements.

\begin{figure}
\centering
\includegraphics[width = 8.8 cm]{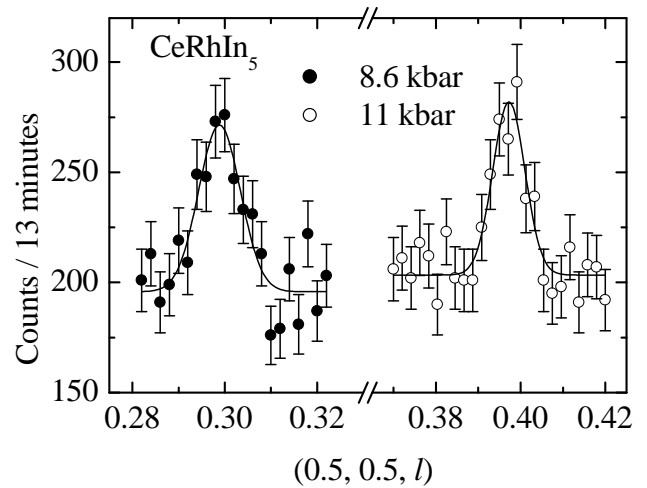}
\caption{ Elastic neutron scans through a magnetic Bragg reflection of
CeRhIn$_5$  at 8.6 kbar  and 11 kbar  at 1.8 K, showing the change in the
value of the magnetic modulation vector with pressure}
\label{fig1}
\end{figure}

\par
\begin{figure}
\includegraphics[width = 8.8 cm]{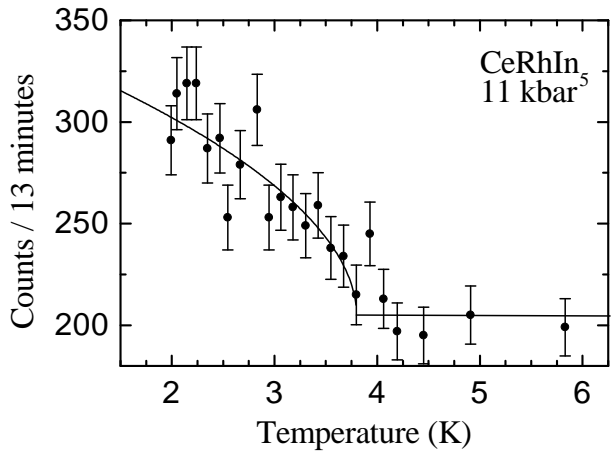}
\caption { Temperature dependence of the magnetic peak (0.5, 0.5, 0.396)
at 11 kbar of pressure. The line through the data points is a guide to the
eyes.}
\label{fig2}
\end{figure}

In order to investigate the nature  of the magnetic phase  at high
pressure, we performed the diffraction study at different hydrostatic
pressures (8.6, 10, 11 and 13 kbars). Before applying each pressure,
the sample was brought back to ambient pressure  and room temperature  to
avoid any irreversibility arising from pressure and  temperature  cycling.
Initially, the crystal was aligned at ambient pressure in the pressure
cell and we observed a magnetic Bragg peak at {\bf q}$_m$ = (0.5, 0.5,
0.2983$\pm$0.0004) at 1.8 K, which is in good agreement with the value
reported by Bao {\it et al}~\cite{neutron1bar}. A hydrostatic pressure of
8.6 kbar was applied  and the crystal was scanned along the (0.5, 0.5,
$\ell$) direction ($\ell$ = 0 to 1) at 1.8 K.  We observed a pair of
magnetic peaks at $\ell$ = 0.2989 and  0.6980 (see Fig. \ref{fig1})
corresponding to a modulation vector {\bf q}$_m$ = (0.5, 0.5, 0.298). We
did not find any other magnetic peaks along this direction. This
magnetic peak vanishes at 3.8 K, showing that the value of $T_N$ remains
unchanged at 8.6 kbar of pressure.  The pressure  was then increased to 11
kbar to observe any change in the magnetic structure. A  scan along (0.5,
0.5, $\ell$) failed to show any magnetic Bragg  peaks around $\ell$ = 0.3
or 0.7 at 1.8 K. Instead, we observed a pair of magnetic peaks at $\ell$ =
0.396 and 0.604 (see Fig. \ref{fig1}). For a spiral magnetic  structure
along the $c$-axis with a nearest neighbor $AFM$ arrangement in the basal
plane, the expected magnetic reflections should occur at ${\bf q}_m =
(h/2, k/2, l\pm \delta)$, where $h, k = \pm$ 1, $\pm$ 3,..., $l =$ 0,
$\pm$ 1, $\pm$ 2,... and $\delta$ is the pitch of the spiral.
Indeed magnetic peaks were also observed at the position (0.5, 0.5,
1.396), (1.5, 1.5, 0.396), (1.5, 1.5, 0.604), (1.5, 1.5, 1.396) and (1.5,
1.5, 1.604) with relative intensities that agree with a model of
basal-plane moments that propagate in a simple helical spiral along the
$c$-direction.  Within our experimental accuracy, no anomalous change in
the position, width or  intensity of the magnetic peak with temperature
was observed at this  pressure.  In Fig. \ref{fig2}, we have plotted the
temperature dependence of the magnetic peak (0.5, 0.5, 0.396) at  11 kbar
of pressure. The peak intensity decreases to the background value above
3.8 K.  No magnetic reflection was seen at the commensurate positions
(0.5, 0.5, 0) or (0.5, 0.5, 0.5). We also scanned the weak (110) nuclear peak
at 1.8  and 5 K to detect any possible signature of ferromagnetic
structure. Within the measurement limit, there was no change in the
intensity of the (110) peak.

\par
In order to obtain  a clearer picture of the functional dependence of {\bf q}$_m$ with
pressure, we have also investigated the magnetic structure at 10 kbar and 13 kbar. The diffraction scan at 10 kbar also showed a  modulation of (0.5,
0.5, 0.396), similar to that observed at 11 kbar. From this observation it
appears that  {\bf q}$_m$ does not change continuously in the pressure
range $9$ kbar $\leq P \leq 11$ kbar, rather it rapidly changes its value
between 8.6 and 10 kbars. It should be noted at
this point,  that the unidentified anomaly in the resistivity {\it vs.}
temperature data appears beyond a pressure value of 9 kbar. The $T_N$
calculated from the heat capacity {\it vs.} temperature data also shows a
downward shift in temperature with an increase of pressure beyond 10 kbar.
Therefore, the pressure at which  we observed the change in  {\bf q}$_m$
matches closely with the pressure values where the  anomalies in
resistivity and heat capacity were seen.

\par
However, at 13 kbar of pressure,  we were unable to detect any magnetic
reflections  in our diffraction scans (at 1.8 K)  along the (0.5, 0.5,
$\ell$) direction. We have also searched for magnetic satellites in
other directions,  {\it e.g.} (1, 1, $\ell$), ($h$, $k$, 0.296), ($h$,
$k$, 0.396), ($h$, $k$, 1), where $\ell$ changes from 0 to 0.5 and $h$ and
$k$ change from 0.45 to 0.55. No magnetic reflection was observed in
these measurements.

\par
Considering the fact that in CeRhIn$_5$ the quasi two-dimensional
CeIn$_3$  building blocks are essential for  the magnetism, it is
pertinent here to make a comparison with the bulk CeIn$_3$ compound. The
compound CeIn$_3$ crystallizes in a cubic structure with a lattice
parameter $a$(300 K) = 4.73 \AA. It orders antiferromagnetically below  10
K with  a {\bf q}$_m$ of (0.5, 0.5, 0.5)~\cite{cein3} and a
superconducting state emerges at 25.5 kbar~\cite{walker} of pressure with
the subsequent quenching of the antiferromagnetic state. The lattice
parameter $a$(300 K) for CeRhIn$_5$ is 4.65 \AA ~\cite{cein3}, which
implies that  the CeIn$_3$ building blocks are in a compressed state  as
compared to the bulk CeIn$_3$ compound. Unlike CeIn$_3$,  CeRhIn$_5$ shows
a spiral magnetic structure. This kind of spiral magnetic structure  is
very common in the rare-earth systems and often occurs due to competing
magnetic interactions~\cite{jensen}. The  major factors responsible for
the magnetic structure in rare-earth systems are: i) the long range
oscillatory exchange interactions of the Rudermann-Kittel-Kasuya-Yosida
($RKKY$) type, ii) the anisotropy energy resulting from crystalline
electric field and iii) the magnetoelastic anisotropy.

\par
 The free electron mediated  $RKKY$ exchange term($\cal{J}$({\bf r}))  is
distance dependent and the sudden change of  $q_m$ in CeRhIn$_5$ above 9
kbar may occur  because of  a change in the interlayer spacing beyond a
critical value($c$(1bar)/$c$(10 kbar) = 1.0052).
Although measured in a high magnetic field,  de Haas van Alphen
experiments in this compound indicate an anisotropic Fermi surface with
some kind of  spin density wave  transition occurring at 1.2 K
~\cite{dhva}. From the Fermi surface point of view, the wave vector
dependent free electron susceptibility, $\chi$($q$) is important in
determining  $\cal{J}$({\bf r}) and for a stable spiral structure
$\chi$($q$) should have  a maximum at a nonzero value of  $q$. It may be
possible that the change in {\bf q}$_m$ is an outcome of the shift in
$\chi$($q$) under pressure resulting from a change in the nesting characteristics of the Fermi surface.  It is also interesting to note that both
$T_N$ and {\bf q}$_m$ are resistant to any change up to a pressure  as
high as 8.6 kbar. The crystal field and the magnetoelastic anisotropy may
not have a direct involvement in changing {\bf q}$_m$, as no change in the
ratio of the lattice parameters $c/a$ was observed at different pressure
values.

\begin{figure}
\centering
\includegraphics[width = 8.8 cm]{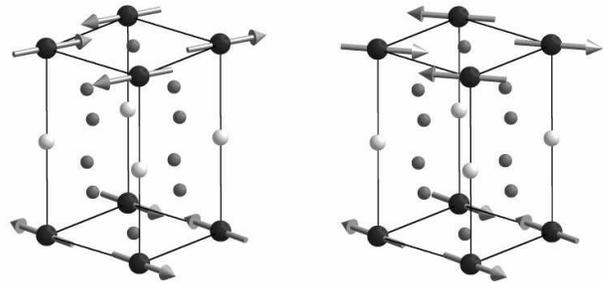}
\caption {An illustration  of the  magnetic structure of CeRhIn$_5$ (a)
below 9 kbar  and  (b) at  11 kbar. The black spheres with spins attached
are the Ce atoms, while the big white spheres and small gray spheres are
Rh and In atoms respectively.}
\label{fig3}
\end{figure}


The  change in {\bf q}$_m$ in the pressure range 8.6 to 10 kbar signifies
a change in the turn angle  of the spiral spin structure from 107$^o$ to
142.6$^o$ (see Fig. 3).
The  effective magnetic moments of CeRhIn$_5$ ($\mu_m$) at  8.6 and 11
kbars were also calculated from the integrated intensity of the magnetic
Bragg peaks. We observed a 20\% reduction of the moment at 11 kbar as
compared to  the 8.6 kbar ($\mu_m$(11 kbar)/$\mu_m$(8.6 kbar) = 0.8 $\pm$
0.1) value. Earlier investigations \cite{neutron3p8kbar} indicate no
change in $\mu_m$ up to a pressure value of  3.8 kbar. The observed change
in $\mu_m$ may be due to the increased Kondo coupling strength along with
the increase of pressure. At 11 kbar we did not observe any anomaly in the
magnetic scattering around  3 K. Therefore, the  unidentified  3 K anomaly
in the  resistivity data~\cite{hegger} above 9 kbar may not be related to
the observed change in the magnetic ordering mechanism. The  heat capacity
measurements~\cite{heatcap}  did not show  any anomaly at 3 K in this
pressure range. Nevertheless, our experiment certainly signifies a change
in the magnetic structure in the same pressure range  where the
unidentified  3 K anomaly develops in resistivity measurements.

\par
 The failure to observe any  magnetic reflection at 13 kbar of pressure
does not necessarily indicate that the system  has attained a non-magnetic
ground state. It might be difficult to observe the magnetic reflection
because :   i) the magnetic moment is reduced below the detectable limit,
ii) $T_N$ has decreased below 1.8 K ,  iii) the magnetic structure  adopts a
completely different  modulation symmetry or  iv) the short range correlations dominate at all temperatures. Interestingly, the pressure dependence
of entropy shows a 2nd order discontinuity around 12 kbar~\cite{heatcap}
and this might be related to the absence of  long range magnetic order  at
13 kbar. The high pressure heat capacity measurements~\cite{heatcap} also
indicate that the magnetic anomaly becomes broader and weaker with
increasing pressure, which might be an indication of the loss of magnetic
moment. Considering CeRhIn$_5$ is a layered compound, a broadening of the
heat capacity peak may also indicate that only anisotropic short range
order or  two dimensional magnetic correlations  exist in the high
pressure state of the system.

\par
Summarizing, this high pressure neutron diffraction study of CeRhIn$_5$
certainly supports
the observed result by heat capacity and resistivity measurements that
$T_N$ remains unchanged up to a pressure value of 10 kbar. In addition, it
indicates a change in the magnetic modulation vector in the pressure range
9-10 kbar. But, in contrast to the results of  heat capacity or
resistivity data, this experiment fails to find any ordered magnetic state at
13 kbar of pressure above 1.8 K. The nature of the magnetism in the superconducting
state is of interest and a neutron scattering  experiment covering a
wider pressure range (above $P_c$) is required to understand the coexistence of magnetism with superconductivity.  Such
an experiment is planned for the near future.

\par
We thank Mr. Jean-Luc Laborier (ILL) for his technical help with the
pressure cell. We also gratefully acknowledge the support from EPSRC for
this project.

\end{document}